\documentclass[review]{elsarticle}

\usepackage{lineno,hyperref}
\usepackage{breakurl}
\usepackage{amsmath,amssymb}
\usepackage{xcolor}
\usepackage{listings}
\usepackage{verbatim}
\modulolinenumbers[5]

\newcommand{\hhat}[1]{\hat{\hat{#1}}}
\usepackage{verbatim}
\newcommand{\new}[1]{{#1}}
\newcommand{\remove}[1]{}
\journal{Computers \& Mathematics with Applications}

\begin{document}

\begin{frontmatter}

\title{GPU acceleration of \emph{CaNS} for massively-parallel direct numerical simulations of canonical fluid flows}

\author[kth,hi]{Pedro Costa\corref{cor}}
\ead{p.simoes.costa@gmail.com}
\cortext[cor]{Corresponding author}
\author[nvd]{Everett Phillips}
\ead{ephillips@nvidia.com}
\author[kth]{Luca Brandt}
\ead{luca@mech.kth.se}
\author[nvd]{Massimiliano Fatica}
\ead{mfatica@nvidia.com}
\address[kth]{Linn\'e FLOW Centre and SeRC (Swedish e-Science Research Centre), KTH Mechanics, SE-100 44 Stockholm, Sweden}
\address[nvd]{NVIDIA Corporation, Santa Clara CA 95050}
\address[hi]{Faculty of Industrial Engineering, Mechanical Engineering and Computer Science, University of Iceland, Hjardarhagi 2-6, 107 Reykjavik, Iceland}

\begin{abstract}
    This work presents the GPU acceleration of the open-source code \emph{CaNS} for very fast massively-parallel simulations of canonical fluid flows. The distinct feature of the many-CPU \new{Navier-Stokes} solver in \emph{CaNS} is its fast \new{direct} solver for the second-order finite-difference Poisson equation, based on the method of eigenfunction expansions. The solver implements all the boundary conditions valid for this type of problems in a unified framework. Here, we extend the solver for GPU-accelerated clusters using CUDA Fortran. The porting \new{makes extensive use of CUF kernels and} has been greatly simplified by the unified memory feature of CUDA Fortran, which handles the data migration between host (CPU) and device (GPU) without defining new arrays in the source code. The overall implementation has been validated against benchmark data for turbulent channel flow and its performance assessed on a NVIDIA DGX-2 system (16 Tesla V100 32Gb, connected with NVLink via NVSwitch). The wall-clock time per time step of the GPU-accelerated implementation is impressively small when compared to its CPU implementation on state-of-the-art many-CPU clusters, as long as the domain partitioning is sufficiently small that the data resides mostly on the GPUs. The implementation has been made freely available and open-source under the terms of an MIT license. 
\end{abstract}


\end{frontmatter}

%\linenumbers

\section{Introduction}
Fluid flows are ubiquitous in nature and industry. Very often these flows are turbulent, exhibiting highly unsteady, chaotic, three-dimensional and multi-scale dynamics. Consistently, the Navier-Stokes equations governing the dynamics of incompressible, Newtonian fluid flows are highly non-linear, which makes analytical predictions often difficult. This challenge, together with the increasing computing power and development of efficient numerical methods has been driving the ever-expanding field of computational fluid dynamics (CFD), which aims to unveil the physics of these complex systems by numerical computations. In particular, Direct Numerical Simulations (DNS) of turbulent flows as a \emph{first-principles} simulation must resolve all spatial and temporal scales of the turbulent flow; one can easily show that the number of operations required for achieving this ambitious goal scales with $\mathrm{Re}_L^3$ \cite{Pope-2000} with $\mathrm{Re}_L$ being a Reynolds number based on the largest flow scales, which can easily reach values of $10^6 - 10^9$ in many real industrial or environmental contexts. By virtue of the aforementioned developments, it is now possible to simulate fluid flows in $O(10^{12})$ spatial degrees of freedom, in relatively simple geometries \cite{Ishihara-et-al-ARFM-2009}, orders of magnitude more than the first DNS of homogeneous isotropic turbulence in the seminal work by \cite{Orszag-and-Patterson-PRL-1972}. Though encouraging, these numbers are still orders of magnitude lower than those required in many real applications.\par
Due to the inherently large computational demand of these simulations (both in terms of memory and processing power), parallel computers based on many Central Processing Units (CPU) have been the machine of choice to tackle DNS of turbulent fluid flows. In the past ten years, however, there has been a paradigm shift in high-end supercomputer architectures, with a strong focus on accelerated computations, in particular with Graphics Processing Units (GPU). The Top500 list of the most powerful supercomputers in the world has been dominated by accelerator-based systems for several years, and at present the current \#1 and \#2 systems in the world are both GPU accelerated \cite{top500}. Accelerated systems (in particular GPU-based systems) are also very power-efficient: the Green500 list of the most efficient supercomputers has also been dominated by accelerated systems in the same time window. Another exciting consequence of this paradigm shift is the increasingly easier access to petascale computing through GPU-based machines, such as the NVIDIA DGX-2 system.\par
GPUs are one of the most popular accelerators. These devices offer high computational power (the Tesla V100 has around $7$ teraflops of 64-bit floating-point peak performance) and high memory bandwidth (the Tesla V100 can sustain around $840$ GB/s on the STREAM benchmark), coupled with the availability of high-level programming languages, numerical libraries and performance/debugger tools. 
GPUs are well-suited  for problems where the arithmetic intensity (the ratio between the floating-point operations performed relative to the amount of memory accessed)  is low, as in finite-difference operations often performed in DNS solvers. The challenge for a many-GPU incompressible DNS solver is parallelizing the remaining tasks that are serial in nature. These are mostly associated with the solution of linear systems of equations for e.g.\ imposing mass conservation, or integrating implicitly in time the diffusion terms in the momentum conservation equation. Fortunately, recent efficient libraries have become available for efficient computations of linear algebra and Fast Fourier Transforms (FFT) on GPUs; e.g.\ in the NVIDIA CUDA Toolkit, or the MAGMA library for linear algebra \cite{MAGMA}.\par
Not surprisingly, numerous recent studies have been devoted towards porting finite-difference DNS codes for incompressible flows in GPU-based architectures. Some examples are the \emph{AFiD} code for wall-bounded turbulent flows with thermal convection \cite{AFID-GPU}; the boundary layer code in \cite{Ha-et-al-JCP-2018,Ha-et-al-arXiv-2018}; and the spectral/finite-difference channel flow code in \cite{Alfonsi-et-al-CaF-2014}; see also the review of CFD calculations on GPUs in \cite{Niemeyer-et-al-JS-2014}. A common outcome in all these studies is the achievement of remarkable computational performance of the GPU implementations, compared to the many-CPU codes used as starting point.\par
The present work describes the extension of a fast DNS solver for massively-parallel calculations on GPU-accelerated clusters. The starting point for this work is the efficient and fast open-source code for  DNS of canonical flows, \emph{CaNS}, described in \cite{Costa-CAMWA-2018}. The solver uses a fast (FFT-based) second-order, finite-difference pressure-correction scheme, where the pressure Poisson equation is solved with the method of eigenfunction expansions. The algorithm explores all combinations of pressure boundary conditions valid for such a solver, in a single and general framework. The method is implemented in Fortran90/95 and extended with a hybrid MPI-OpenMP parallelization, with a 2D pencil-like domain decomposition, which enables efficient massively-parallel simulations. Several recent examples of numerical implementations using this direct solver, combined with a 2D domain decomposition, achieved unprecedented performances for complex flows in domains with $O(10^{9})-O(10^{10})$ grid points, see e.g.\ \cite{Costa-et-al-PRL-2016,Dodd-and-Ferrante-JFM-2016,Ostilla-et-al-JFM-2016}. Despite the complexity of the systems addressed in these references, the efficient base Navier-Stokes solver used is a key element that has made the simulations therein presented in reach.\par
Here, we extend \emph{CaNS} for computations on GPUs using CUDA Fortran. With its recent unified (or managed) memory feature, we were able to port the code to GPU architectures, mostly with small changes in the original source, while still reaching excellent computational performance. The GPU extension has been validated and its performance assessed. The results show a code performance on $4$ NVIDIA Tesla V100 GPUs on a DGX-2 to be \new{about the same ($0.9$ times slower) to $1.6$ times faster} as the CPU code on \new{$2048$} cores on state-of-the-art CPU-based supercomputers, and \new{$3.1$ to $5.6$} times faster when all the $16$ GPUs of the DGX-2 cluster are used. \remove{Among the authors, PC developed the original implementation of \emph{CaNS} with support from LB, while EP and MF developed the GPU extension.}\par
This paper is organized as follows. Next, section~\ref{sec:numerics} describes the overall numerical method and the approach used in the fast Poisson solver. After, section~\ref{sec:implementation} summarizes the many-CPU implementation in \emph{CaNS} and presents in detail the approach for the many-GPU extension. Section~\ref{sec:results} validates the implementation and presents the computational performance of the many-GPU extension. Finally, in section~\ref{sec:conclusions} we summarize the main conclusions of the work\remove{ and present future perspectives}.

\section{Numerical method}\label{sec:numerics}
The numerical algorithm solves the Navier-Stokes equations for an incompressible, Newtonian fluid with constant unit density $\rho=1$ and dynamic viscosity $\mu$ (kinematic viscosity $\nu$),
\begin{subequations}
\begin{align}
    \boldsymbol{\nabla} \cdot \mathbf{u} = 0 \mathrm{,} \label{eqn:cont} \\
    \rho\left(\frac{\partial \mathbf{u}}{\partial{t}} + \nabla \cdot (\mathbf{u}\otimes\mathbf{u})\right) = \nabla\cdot{\boldsymbol\sigma}\mathrm{,}
\end{align}
\end{subequations}
where the stress tensor $\boldsymbol{\sigma}=-p\boldsymbol{I}+\mu(\nabla \mathbf{u}+\nabla \mathbf{u}^T)$, with $\mathbf{u}$ and $p$ being the fluid velocity vector and pressure.\par
These equations are solved on a structured Cartesian grid, uniformly-spaced in two directions. The method uses second-order finite-differences for spatial discretization with a staggered (marker and cell) disposition of grid points, and a low-storage, three-step Runge-Kutta scheme for time integration \cite{Wesseling-2009}. For the sake of clarity the numerical scheme is summarized below, and we refer to \cite{Costa-CAMWA-2018} for more details.\par
The advancement at each substep $k$ reads ($k=0,1,2$; $k=0$ corresponds to a time level $n$ and $k=3$ to $n+1$):
{\small\begin{subequations}
 \begin{align}
 \mathbf{u}^* = \mathbf{u}^k + \Delta t\left(\alpha_k\left(\mathcal{A}\mathbf{u}^{k}+\nu\mathcal{L}\mathbf{u}^{k}\right) + \beta_k\left(\mathcal{A}\mathbf{u}^{k-1} + \nu\mathcal{L}\mathbf{u}^{k-1}\right) - \gamma_k\mathcal{G} p^{k-1/2}\right)\mathrm{,} \label{eqn:up} \\
 \mathcal{L}\Phi = \frac{\mathcal{D}\mathbf{u}^*}{\gamma_k\Delta t}\mathrm{,}\label{eqn:poi_ns} \\
 \mathbf{u}^{k+1} = \mathbf{u}^* - \gamma_k\Delta t \mathcal{G}\Phi\mathrm{,} \\
 p^{k+1/2} = p^{k-1/2} + \Phi\mathrm{,}
 \end{align}\end{subequations}}
where $\mathcal{A}$, $\mathcal{L}$, $\mathcal{G}$, and $\mathcal{D}$ denote the discrete advection, Laplacian, gradient and divergence operators; $\mathbf{u}^*$ is the prediction velocity and $\Phi$ the correction pressure. The RK3 coefficients are given by $\alpha=\lbrace{8/15,5/12,3/4\rbrace}$, $\beta=\lbrace{0,-17/60,-5/12\rbrace}$ and $\gamma=\alpha+\beta$. For low Reynolds number flows, or very fine grids, implicit temporal discretization of the diffusion term can be desirable. In this case, the temporal integration is as follows:
{\small\begin{subequations}
 \begin{align}
 \mathbf{u}^{**} = \mathbf{u}^k + \Delta t\left(\alpha_k\mathcal{A}\mathbf{u}^{k} + \beta_k\mathcal{A}\mathbf{u}^{k-1} + \gamma_k\left(-\mathcal{G} p^{k-1/2}+\nu\mathcal{L}\mathbf{u}^{k}\right)\right)\mathrm{,} \label{eqn:up_id} \\
 \mathbf{u}^{*}-\gamma_k\frac{\nu\Delta t}{2}\mathcal{L}\mathbf{u}^{*} = \mathbf{u}^{**} - \gamma_{k}\frac{\nu\Delta t}{2} \mathcal{L}\mathbf{u}^{k}\label{eqn:up_id_2}\\
 \mathcal{L}\Phi = \frac{\mathcal{D}\mathbf{u}^*}{\gamma_k\Delta t}\mathrm{,}\label{eqn:poi_ns_id} \\
 \mathbf{u}^{k+1} = \mathbf{u}^* - \gamma_k\Delta t \mathcal{G}\Phi\mathrm{,} \\
p^{k+1/2} = p^{k-1/2} + \Phi - \gamma_k\frac{\nu\Delta t}{2}\mathcal{L} \Phi\mathrm{.}
\end{align}\end{subequations}}
Note that eqs.~\eqref{eqn:up_id} and~\eqref{eqn:up_id_2} are not combined, to illustrate that $\mathbf{u}^{**}$ is a better approximation of the final velocity than the sum of the terms on the right-hand-side of eq.~\eqref{eqn:up_id_2}. This splitting is desirable e.g.\ in case of direct forcing immersed-boundary methods (IBM), for $\mathbf{u}^{**}$ is the prediction velocity from which the IBM force should be computed; see e.g.\ \cite{Uhlmann-JCP-2005}.\par
%
%brief description of the numerical method
A sufficient criterion for a stable temporal integration is given in \cite{Wesseling-2009}:
 \begin{equation}
 \Delta t < \min\left(\frac{1.65\Delta \ell^2}{\nu},\frac{\sqrt{3}\Delta \ell}{\max_{ijk}(|u|+|v|+|w|)}\right)\mathrm{,}\label{eqn:dt}
 \end{equation}
 with $\Delta \ell = \min(\Delta x,\Delta y,\Delta z)$ and $\Delta {x_i}$ the grid spacing in direction $x_i=\lbrace{x,y,z\rbrace}$, where the time step restriction due to the viscous effects is absent with implicit treatment of the diffusion term (left term on the right-hand-side of eq.~\eqref{eqn:dt}).

\subsection*{FFT-based Poisson solver}
Since the solution of the Poisson/Helmholtz equations introduced above comprises the most elaborate implementation steps, these are summarized below. The Poisson equations (eqs.~\ref{eqn:poi_ns} and~\ref{eqn:poi_ns_id}) are discretized in space \new{at grid cell $i,j,k$} as follows:
 \begin{align}
   &(\Phi_{i-1,j  ,k  }-2\Phi_{i,j,k}+\Phi_{i+1,j  ,k  })/\Delta x^{2} + \nonumber\\
   &(\Phi_{i  ,j-1,k  }-2\Phi_{i,j,k}+\Phi_{i  ,j+1,k  })/\Delta y^{2} + \nonumber\\
   &(\Phi_{i  ,j  ,k-1}-2\Phi_{i,j,k}+\Phi_{i  ,j  ,k+1})/\Delta z^{2} = f_{i,j,k}\mathrm{;}\label{eqn:poi}
 \end{align}
\new{with $f_{i,j,k}$ being the right-hand-side of eq.~\eqref{eqn:poi_ns} or~\eqref{eqn:poi_ns_id}  at grid cell $i,j,k$.} The solution method reduces this system of equations, with $7$ non-zero diagonal terms to a tridiagonal system, which can be solved very efficiently with Gauss elimination. To achieve this we apply the discrete operator $\mathcal{F}^{x_i}$ to eq.~\eqref{eqn:poi} in two domain directions, which reduces the problem to:
 \begin{align}
   (\lambda_{i}/\Delta x^2 + \lambda_{j} /\Delta y^2)\hhat{\Phi}_{i,j,k} +
   (\hhat{\Phi}_{i,j,k-1}-2\hhat{\Phi}_{i,j,k}+\hhat{\Phi}_{i,j,k+1})/\Delta z^{2} = \hhat{f}_{i,j,k}\mathrm{,} \label{eqn:poi_reduced}
 \end{align}
 where $\hhat{\square}=\mathcal{F}^y(\mathcal{F}^x(\square))$. The discrete operator $\mathcal{F}^{x_i}$ can be expressed in terms of discrete Fourier transforms and depends on the problem's boundary conditions; see \cite{Costa-CAMWA-2018,Schumann-and-Sweet-JCP-1988} for more details. We should note that the equations are written assuming a uniform grid spacing in $z$ for simplicity; as we will show, the grid in $z$ can be non-uniform.\par
Finally, in the case of implicit treatment of viscous momentum diffusion, the three Helmholtz equations in eq.~\eqref{eqn:up_id_2} are solved with the same type of direct solver, rather than following the (computationally cheaper) alternating direction implicit approach used e.g.\ by Kim \& Moin~\cite{Kim-and-Moin-JCP-1985}, where a third-order-in-time approximation of the system of equations can be solved with three sequential tridiagonal solves per velocity component. We preferred using the same direct FFT-based solver as that of the pressure, as we found it more straightforward to generalize the approach to different combinations of boundary conditions.
\section{Implementation}\label{sec:implementation}
%%\footnote{Open source and freely available in \burl{github.com/p-costa/CaNS} under the terms of an MIT license.}
\subsection*{Many-CPU implementation with MPI-OpenMP in \emph{CaNS}}
    Here we summarize the original implementation of \emph{CaNS} for massively-parallel DNS in CPU clusters and refer to \cite{Costa-CAMWA-2018} for more details. As mentioned above, the numerical algorithm is implemented in Fortran90/95, extended with MPI-OpenMP for distributed-memory parallelization. The domain is partitioned into several computational subdomains in a 2D \emph{pencil}-like decomposition. In most steps of the calculation, the domain is partitioned in $x$ and $y$ into
    $N_p^x\times N_p^y$ pencils, aligned in the $z$ direction. The $2$DECOMP$ \& $FFT library is used for performing the data transpositions to $x$- and $y$-aligned pencils, which are required for computing the FFT-based transforms. The vectors of real-to-real FFT-based transforms are computed using the GURU interface of the FFTW library \cite{Frigo-and-Johnson-1998}. A very convenient feature of this interface is that each of the $9$ types of fast discrete transforms that are used (dictated by
    the boundary conditions of the Poisson/Helmholtz equation) are computed with exactly the same syntax, just by evoking the right transform type in the \emph{planner}, and considering the different scaling factors.

%\new{Contrarily to the CPU version, the current GPU extension only allows for periodic boundary conditions in the two regular directions, because unlike FFTW, the cuFFT library does not explicitly perform all real-to-real transforms that can be exploited for imposing different boundary conditions. The next step is therefore to implement the different real-to-real transforms by pre- and post-processing FFTs \cite{Makhoul-1980,Schumann-and-Sweet-JCP-1988,Hasbestan-and-Senocak-arXiv-2019}.}\par
%
\subsection*{Many-GPU extension with MPI-CUDA Fortran}
For this specific porting effort, we used CUDA Fortran \cite{CUDAFortranBook} since the original CPU code is in Fortran90/95.
CUDA Fortran is an analog to the NVIDIA CUDA C compiler. It provides both a lower-level explicit programming model that gives direct access to all aspects of GPU programming and a higher-level implicit programming  model via kernel loop directives (CUF kernels).  It is similar to what OpenACC offers but simpler, since CUF kernels can only be applied to nested loops and scalar reductions and all the data movements/allocations is left to the programmer. They are nevertheless a very powerful tool; indeed, most of all the porting was done with CUF kernels. In order to use CUF kernels, the PGI compiler needs to be used (the IBM XLF compiler is able to compile explicit CUDA Fortran but does not support CUF kernels). PGI offers a freely available community edition of their compiler on both x86 and Power systems (support for ARM systems has just been announced), so this restriction is not an issue.\par
In a typical GPU-accelerated code, since the CPU and GPU have different memory spaces, for each array defined on the CPU (host) there will be an equivalent array defined on the GPU (device). A consistent memory view will be enforced by the programmer, copying data back and forth between host and device before operating on them. All the array declarations  need to be duplicated and explicit copies need to be inserted in the code.
A typical sequence of operations for CUDA Fortran code will look like:
\begin{itemize}
\item Declare and allocate host and device memory;
\item Initialize host data;
\item Transfer data from the host to the device;
\item Execute one or more kernels;
\item Transfer results from the device to the host.
\end{itemize}
New features in Fortran (like molded and sourced allocations) may help to reduce the amount of code that needs to be added.
This was the approach used in a previous porting of  a similar DNS code \cite{AFID-GPU}, where all the arrays were explicitly re-declared in  device memory. This porting is instead  using a recent feature called unified (or managed) memory, which dramatically simplifies GPU programming, making arrays accessible from either the GPU or the CPU.
With managed memory the previous sequence of operations will look like:
\begin{itemize}
\item Declare and allocate managed  memory;
\item Initialize  data;
\item Execute one or more kernels.
\end{itemize}
With managed memory, the data movement still occurs, but, rather than being explicit, it is now controlled by the unified memory management system behind the scenes, similarly to the operating system managing virtual memory.
DNS are very amenable to this approach: after the initialization or restart, the flow field will reside in GPU memory essentially all the time. Since a simulation will run for several thousands iterations, the initialization part is usually a negligible portion  of  the total runtime. There are also hints and prefetch commands that can be given to the compiler to optimize the data traffic.
With managed memory, the GPU memory can be in principle over-subscribed: while the code will run and give the correct result, the speed of execution will be severely impacted since the data will continually migrate over the PCI-e bus (with a typical transfer speed of $10$ GB/s) on x86 systems or NVlink (with a typical transfer speed of $30-50$ GB/s) on Power system, an order of magnitude smaller than on device memory. %There are also some effects on GPU-aware MPI that we will describe later.
\par
In order  to have a code as close as possible to the original CPU version, the GPU implementation makes extensive use of the preprocessor, and all the GPU specific code and directives are guarded by {\em USE\_CUDA} macro or sentinel {\em !@cuf} (this is similar to the {\em !$omp$} sentinel defined only when OpenMP is enabled, in this case the sentinel is active when the compiler generates code for the GPU).
For the same Fortran90 source file, a CPU object file can be created with the standard optimization flags  while a GPU version can be created adding the  "\texttt{-DUSE\_CUDA -Mcuda}" flags.

A typical subroutine will look like listing~1.
When compiled for the GPU, line 2 imports the module cudafor to access the \texttt{cudaDeviceSynchronize()} routine (this is needed to ensure that the \texttt{mpi\_allreduce} call is executed only after the kernel, since a typical kernel will run asynchronously).
The \texttt{ifdef} macro at line 7 adds the managed attribute to the arrays that are accessed in the kernel.
The \texttt{ifdef} macro at line 15 generates the GPU kernel for the triple nested loop following, or alternatively 
generate the multithreaded code for CPU with OpenMP. The cuf kernel directive is very simple to use; there is no need to indicate that \texttt{dti} is a reduction variable.\par
\begin{minipage}{\linewidth}
\begin{lstlisting}[caption={Source code for the computation of the maximum allowable time step, $\Delta t$.},captionpos=b]
 subroutine chkdt(n,dl,dzci,dzfi,visc,u,v,w,dtmax)
    !@cuf use cudafor
    implicit none
    ...
    real(rp), intent(in), dimension(0:) :: dzci,dzfi
    real(rp), intent(in), dimension(0:,0:,0:) :: u,v,w
#ifdef USE_CUDA
    attributes(managed):: u,v,w,dzci,dzfi
    integer:: istat
#endif
    integer :: i,j,k
    !
    dti = 0.
    ...
#ifdef USE_CUDA
    !$cuf kernel do(3) <<<*,*>>>
#else
    !$OMP PARALLEL DO DEFAULT(none) &
    !$OMP SHARED(n,u,v,w,dxi,dyi,dzi,dzci,dzfi) &
    !$OMP PRIVATE(i,j,k,ux,uy,uz,vx,vy,vz,wx,wy,wz,dtix,dtiy,dtiz) &
    !$OMP REDUCTION(max:dti)
#endif
    do k=1,n(3)
      do j=1,n(2)
        do i=1,n(1)
          ux = abs(u(i,j,k))
          vx = 0.25*abs( v(i,j,k)+v(i,j-1,k)+v(i+1,j,k)+v(i+1,j-1,k) )
          wx = 0.25*abs( w(i,j,k)+w(i,j,k-1)+w(i+1,j,k)+w(i+1,j,k-1) )
          dtix = ux*dxi+vx*dyi+wx*dzfi(k)
          ....
          dtiz = uz*dxi+vz*dyi+wz*dzci(k)
          dti = max(dti,dtix,dtiy,dtiz)
        enddo
      enddo
    enddo
#ifndef USE_CUDA
    !$OMP END PARALLEL DO
#endif
!@cuf istat=cudaDeviceSynchronize()
    call mpi_allreduce(MPI_IN_PLACE,dti,1,MPI_REAL_RP,MPI_MAX,MPI_COMM_WORLD,ierr)
    ...
    return
  end subroutine chkdt
\end{lstlisting}
\end{minipage}
\subsection{CUF kernels}
CUDA Fortran allows automatic kernel generation and invocation from a region of host code containing one or more tightly nested loops.
Launch configuration and mapping of the loop iterations onto the hardware is controlled and specified as part of the directive body using the familiar CUDA chevron syntax: the developer can specify a particular launch configuration or delegate the choice to the compiler. As with any kernel, the launch is asynchronous and the program can use \texttt{cudaDeviceSynchronize()} or CUDA Events to wait for the completion of the kernel.
The work in the loops specified by the directive is executed in parallel, across the thread blocks and grid.
CUF kernels can also handle  scalar reduction operations, such as summing the values in a vector or matrix. For these operations, the compiler handles the generation of the final reduction kernel, inserting synchronization into the kernel as appropriate.
\subsection{\new{Implementation of the FFT-based transforms using cuFFT}}
\new{The FFTs required by the Poisson/Helmholtz solver are computed with the cuFFT library from the CUDA Toolkit, and are parallelized with the same approach as in the AFiD code (see \cite{AFID-GPU} for details). Despite having a similar interface to the FFTW library used in the CPU version, cuFFT does not support the family of real-to-real transforms implemented in FFTW. Therefore, the different real-to-real transforms must be implemented by pre- and post-processing FFTs
\cite{Makhoul-1980,Schumann-and-Sweet-JCP-1988,Hasbestan-and-Senocak-arXiv-2019}. Some of these transforms have been implemented in the GPU version, namely the standard fast discrete sine and cosine transforms DCT-II and DST-II (and the corresponding inverse transforms, DCT-III and DST-III), following the low-storage approach of Makhoul~\cite{Makhoul-1980}. Hence, all the flows presented in the paper describing the CPU version \cite{Costa-CAMWA-2018} -- a lid-driven cavity,
a pressure-driven turbulent channel and square duct, and a decaying Taylor-Green vortex -- can be also simulated with the GPU version.}
\subsection{Profiling using NVTX} \label{Profiling}
%%%%%%%%%%%%%%%%%%%%%%%%%%%%%%%%%%%%
% NVTX Figures %
%%%%%%%%%%%%%%%%%%%%%%%%%%%%%%%%%%%%
\begin{figure*}
\begin{centering}
\includegraphics[width=0.95\textwidth]{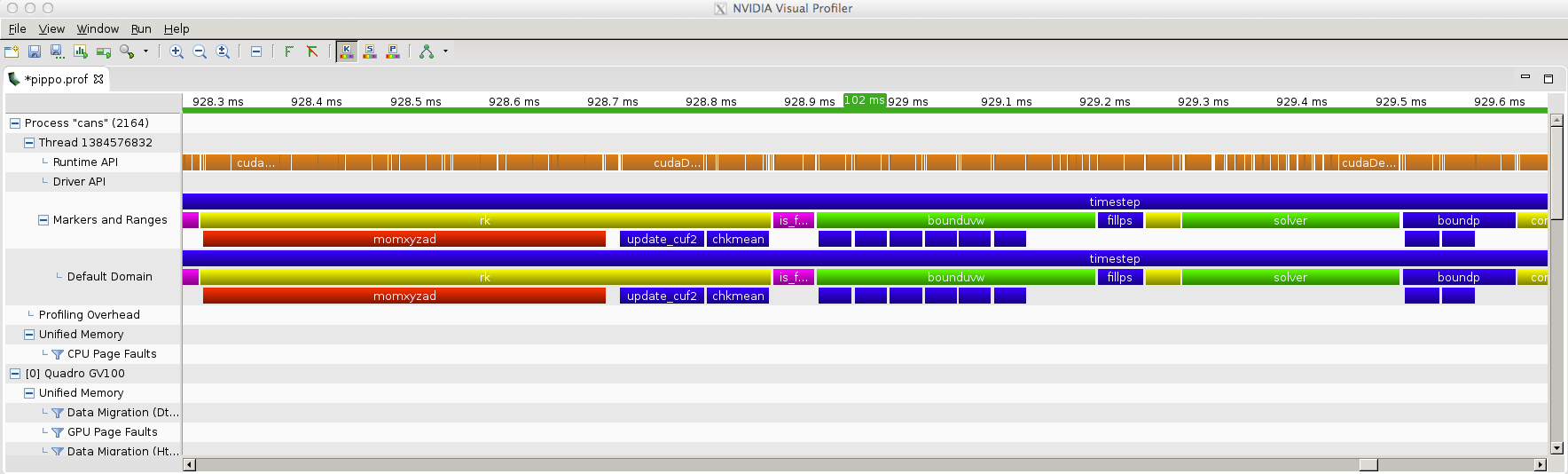}
\caption{Example nvvp session with markers from NVTX.} \label{fig:nvvp}
\end{centering}
\end{figure*}
Profiling is an essential part of performance tuning. It allows to identify parts of the code that may require additional attention. When dealing with GPU codes, profiling is even more important as new opportunities for better interactions between the CPUs and the GPUs can be discovered. The standard profiling tools in CUDA (nvprof and nvvp or the new NSight tools), are able to show the GPU timeline but do not present CPU activity. The NVIDIA Tools Extension (NVTX) is a C-based library that can annotate the profiler time line with events and ranges, can customize their appearance and can assign names to resources such as CPU threads and devices \cite{NVTX}.\par
We have written a Fortran module to instrument CUDA/OpenACC Fortran codes using the Fortran ISO C bindings \cite{NVTXFortran}. The use is very simple, once the NVTX module is loaded, the developer needs to mark the region of interest with \texttt{{nvtxRangePush}} and \texttt{nvtxRangePop} calls. Calls to \texttt{ nvtxStartRange("text")} with a single argument will insert green markers with a \texttt{text} label in the timeline. Different colors can be selected using an optional integer parameter and the regions of interest can be nested. Fig.\ \ref{fig:nvvp} shows an example for \emph{CaNS}.
\section{Validation and Computational Performance}\label{sec:results}
A turbulent channel flow at friction Reynolds number $\mathrm{Re}_\tau\approx 590$ \cite{Moser99} was simulated to validate the code and assess performances. The flow is driven by a uniform pressure gradient that ensures a constant bulk velocity $U_b$. The simulation has been carried in a computational domain with parameters $N_x/L_x\times N_y/L_y\times N_z/L_z=1536/(6h)\times768/(3h)\times576/(2h)$, where $N/L$ denotes the number of grid points/domain length, $h$ is the channel half height, and the subscripts $x$, $y$ and $z$ denote the streamwise, spanwise and wall-normal directions. The grid is regular in $x$ and $y$, as per requirement of the fast Poisson solver, and periodic boundary conditions are imposed therein. In the wall-normal direction the grid is non-uniform, clustered at the two walls. Following \cite{Orlandi-2012}, the centered wall-normal position of a grid cell $i,j,k$ is given by
\begin{equation}
z_k = \frac{1}{2}\left(1+\frac{\tanh\left[a\left(Z_k-0.5\right)\right]}{\tanh{\left[a\left(1-0.5\right)\right]}}\right)L_z\mathrm{,}
\end{equation}
with $Z_k = (k-0.5)/N_z,\,\, k=1,2,\dots,N_z$; the grid clustering parameter is set to $a=1.6$, so that a resolution of about one viscous wall unit is achieved near each wall. No-slip and no-penetration boundary conditions are imposed at the  walls, i.e.\ at $z=h\mp h$.
We set the bulk Reynolds number $\mathrm{Re}_b=U_b(2h)/\nu=12700$, estimated from $\mathrm{Re}_\tau=0.09\mathrm{Re}_b^{0.88}$ to yield the desired pressure drop \cite{Pope-2000}. We expect a corresponding friction Reynolds number close to the target value, but not exactly $\mathrm{Re}_\tau\approx 590$, due to the uncertainty of the correlation. The flow is initialized with a laminar Poiseuille velocity field, together with a high amplitude disturbance consisting of streamwise counter-rotating vortices, to trigger turbulence effectively \cite{Henningson-and-Kim-JFM-1991}. The simulation has been carried out with explicit temporal integration of the diffusion term, as the maximum allowed time step in this problem is dictated by advection (i.e.\ second term on the right-hand-side of eq.~\eqref{eqn:dt}). The system was simulated for $300\,000$ time steps, corresponding to a total physical time in bulk units of about $600(2h)/U_b$.\par
\begin{figure}[!t]
 \centering
   \includegraphics[width=0.99\textwidth]{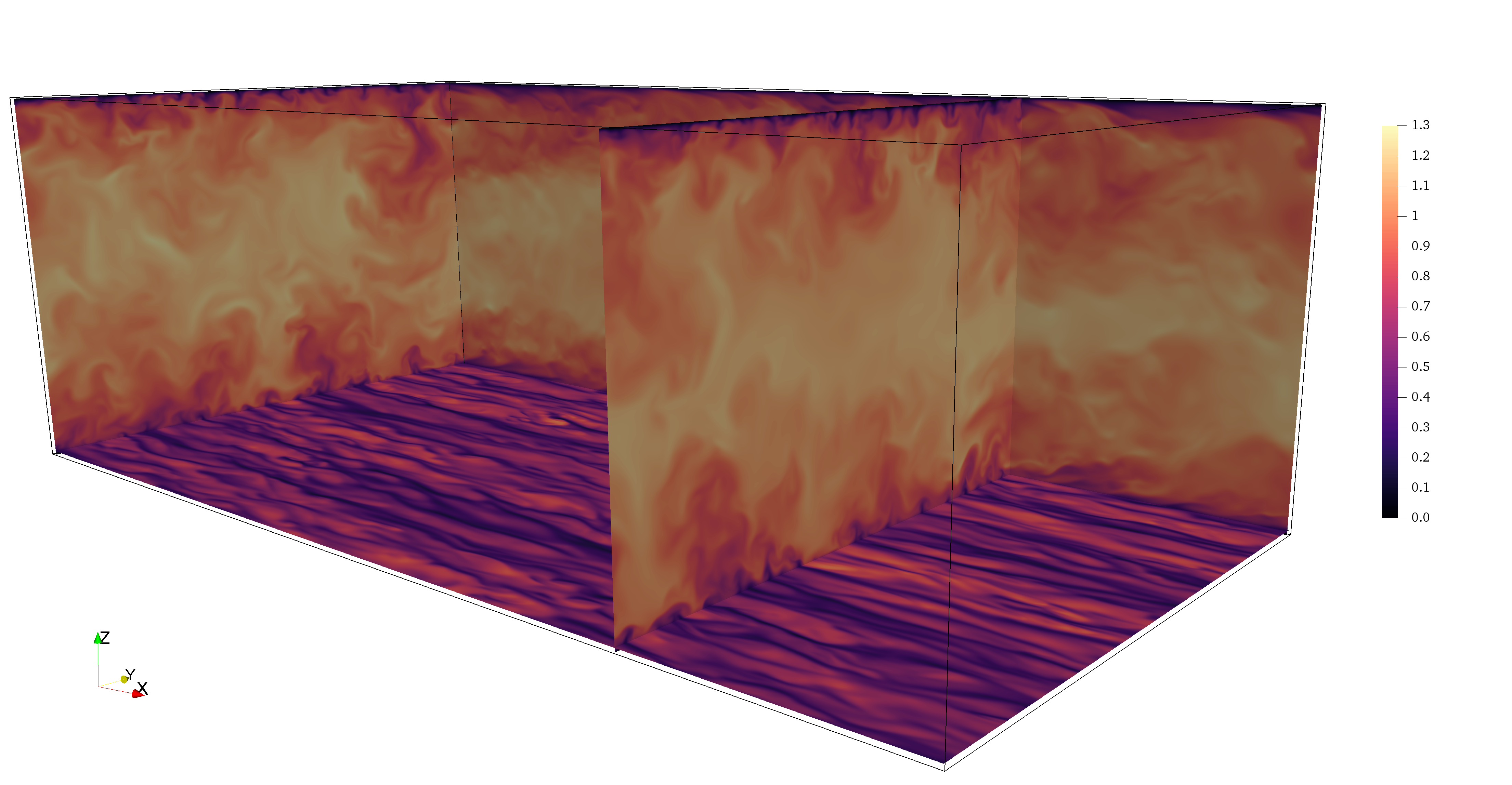}
   \caption{Visualization of the turbulent channel flow. The planes show the contours of streamwise velocity. The wall-parallel plane is located at $z^+=z u_\tau/\nu\approx 12$.}\label{fig:visu}
 \end{figure}
\begin{figure}[!htbp]
 \centering
   \includegraphics[width=0.65\textwidth]{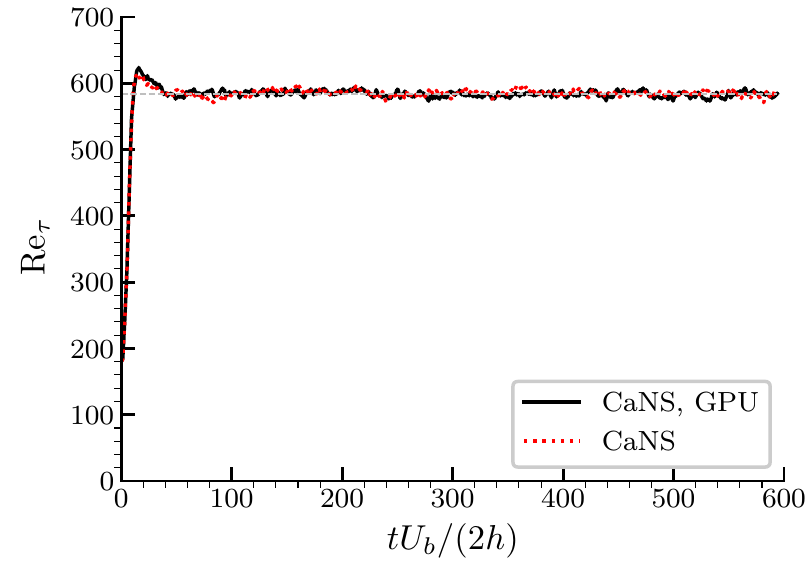}
   \caption{Time evolution of the pressure drop, expressed in terms of the friction Reynolds number $\mathrm{Re}_\tau$. The gray dashed line corresponds to the time average in the fully-developed regime $\mathrm{Re}_\tau = 583.8$.}\label{fig:valchan_presgrad}
 \end{figure}

Figure~\ref{fig:visu} depicts a three-dimensional visualization of the flow. The planes showing contours of streamwise velocity clearly illustrate some of the usual features of turbulent channel flow, such as near-wall low- and high-speed streaks.
The evolution of the pressure drop is shown in figure~\ref{fig:valchan_presgrad}, expressed in terms of the friction Reynolds number $\mathrm{Re}_\tau=u_\tau h/\nu$ with $u_\tau=\sqrt{(-\mathrm{d}p/\mathrm{d}x)h}$. The initial condition effectively triggers transition, and the flow reaches a fully-developed state at $t\approx 100 (2h)/U_b$, when the friction Reynolds number fluctuates around the mean value of $\mathrm{Re}_\tau = 583.8$. The dashed red line in the same figure shows same quantity computed from the CPU implementation. It can be easily seen that, as transition is triggered, the results from the GPU and CPU start to fluctuate around the same mean pressure drop, with different instantaneous values. This is attributed to the chaotic nature of the governing equations, which are extremely sensitive to round-off errors. In particular, the codes have been compiled in different systems, using different FFT libraries, which is likely the cause for the deviations. Expectedly, the time-averaged statistics are the same for both CPU and GPU simulations.
 \par
\begin{figure}[!htbp]
 \centering
   \includegraphics[width=0.65\textwidth]{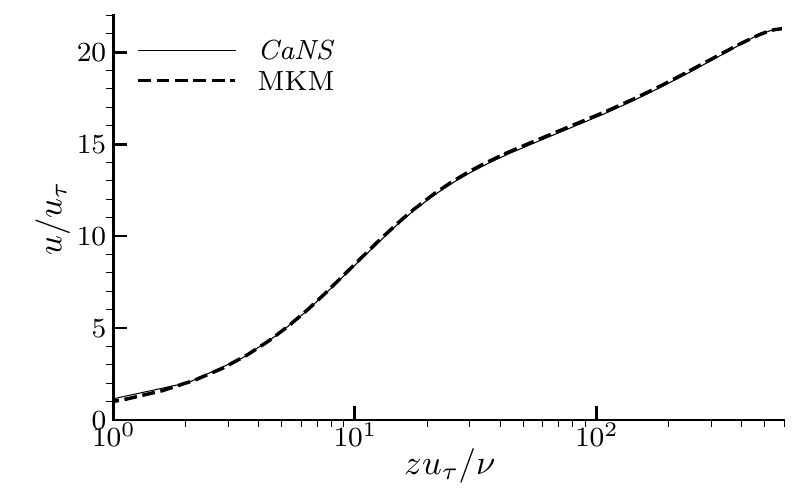}
   \put(-220,140){(\textit{a})}\\
   \includegraphics[width=0.65\textwidth]{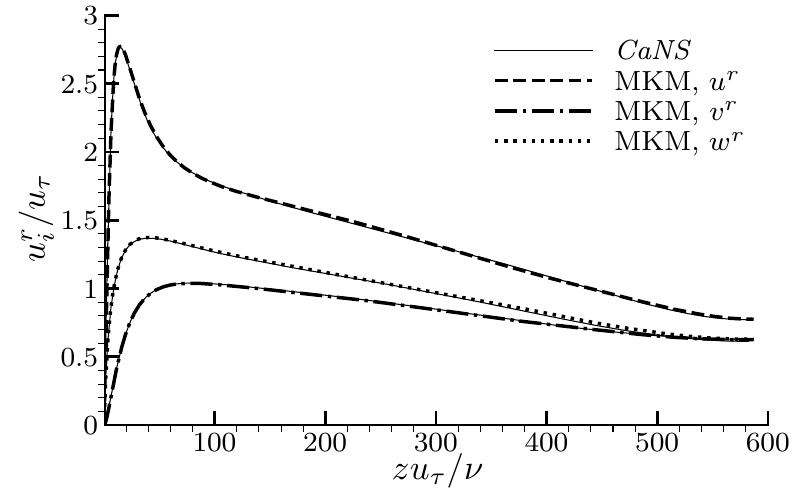}
   %\put(-345,100){(\textit{a})}
   %\put(-170,100){(\textit{b})}
   \put(-220,140){(\textit{b})}

   \caption{(\textit{a}): mean streamwise velocity profile for turbulent channel flow at friction Reynolds number $\mathrm{Re}_\tau\approx 590$. (\textit{b}): profiles of root-mean-square velocity $u_i^r$. Both figures use     inner-scaling, i.e.\ velocity scaled with the wall friction velocity $u_\tau$, and distance with the viscous wall-unit $\nu/u_\tau$. The profiles are compared to DNS data from \cite{Moser99} (MKM).}\label{fig:valchan_vel}
 \end{figure}
\begin{figure}[!htbp]
 \centering
   \includegraphics[width=0.65\textwidth]{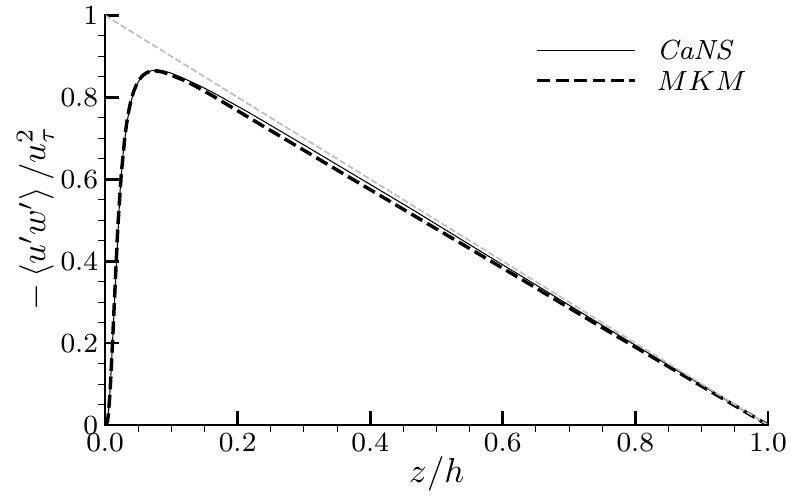} \hfill
   \caption{Profile of inner-scaled Reynolds shear-stresses $-\left<u^\prime v^\prime\right>$ versus the outer-scaled wall-normal distance. The profiles are compared to DNS data from \cite{Moser99} (MKM). The dashed-gray line corresponds to the analytical profile of total stresses.}\label{fig:valchan_reystr}
\end{figure}

Figure~\ref{fig:valchan_vel} compares the present results to those of \cite{Moser99}, for the inner-scaled profiles of the mean streamwise velocity (panel (\textit{a})), and the three velocity r.m.s.\ (panel (\textit{b})). The present results correspond to ensemble-averages of $300$ samples in the fully-developed regime, equally-spaced over a time interval of about $300\,(2h)/U_b$. The agreement with the reference data is excellent, which validates the GPU implementation. The good agreement also holds for the Reynolds stresses profile shown in figure~\ref{fig:valchan_reystr}, where the minor differences in the bulk of the channel are attributed to the slightly smaller friction Reynolds number in the present simulation.
\par
Next, we assess the performance of the GPU implementation on state-of-the-art GPU-based systems. The simulations with the GPU code have been performed on two systems: a DGX Station with $4$ NVIDIA Tesla V100 32GB, a DGX-2 system with $16$ V100 32 GB.
The wall-clock time per time step for the simulation in the DGX-2 system is presented in table~\ref{tbl:performances}, for different computational grids (i.e.\ distributions of computational subdomains). Due to memory requirements, the DGX Station can run the selected grid only with the implicit diffusion turned off: since the code is using managed memory, the simulation will run but the wall-clock time per time step will be an order of magnitude longer, since there is a lot of memory migration
between the CPU and GPU. The time per time step of the run on the four GPUs of DGX Station is $0.7\mathrm{s}$. \new{Remarkably, when the full DGX-2 system is used, the implementation reaches a very low wall-clock time per time step that allows to reach a fully-developed state ($\approx 38000$ time steps) in $1.5$ hours.} \par
From Table~\ref{tbl:performances}, we can also see that a 1D decomposition of the problem is always faster,
since one of the transpose steps in the all-to-all communication, required by the Poisson solver,
is done in the GPU memory instead of going through NVLink.
When the implicit diffusion is active, the gain is even more pronounced since there are more \textit{all-to-all} communications that need to be performed. The difference in wall-clock time per  step between the 4 GPU runs on the DGX Station ($0.7$s, not reported in Table~\ref{tbl:performances}) and DGX-2 system ($0.48$s, first line in Table~\ref{tbl:performances}) can also be explained by the different bandwidth of NVLink on these two systems. NVLink speed depends on the number of links active. Each Tesla V100 has 6 links available, where each link has a signal rate of $25$ GB/s. In the DGX-2 system, each of the $16$ GPUs has $6$ links active (translating into $150$ GB/s) and is connected to all the others via a NVSwitch that can simultaneously drive communications between all 8 GPU pairs at full speed. On the DGX Station, there is no NVSwitch and the $4$ GPUs are connected to each other directly. Some of the GPUs are connected to each other via a single link, and others via two links.\par
The same setup has been simulated with the CPU version, in a computational grid of \new{$64\times 32$}, i.e.\ \new{$2048$} cores, in two supercomputers based in Sweden: Beskow (151 in the June 2019 TOP500 list; Cray XC40, Xeon E5-2695v4/E5-2698v3 16C 2.3GHz, Aries interconnect), and the more recent Tetralith (74 in the June 2019 TOP500 list; Intel H2204XXLRE, Xeon Gold 6130 16C 2.1GHz, Intel Omni-Path), the timings pertaining to these machines for the same channel flow DNS, with diffusion treated
explicitly, are shown in the caption of table~\ref{tbl:performances}. Quite remarkably, the simulations on the DGX-2 system with $4$ GPUs are \new{about as expensive ($0.9$ times slower) to $1.6$} times faster than the CPU simulations, and \new{$3.1$ to $5.6$} times faster when all the $16$ GPUs of the system are used. \new{Assuming that the speedup of the many-CPU simulation for this setup scales linearly with further increase of the number of cores (which is an extremely conservative premise, since the load per task in this case becomes too small for the CPU implementation speedup to scale linearly), we can estimate that the CPU system requires $6100$ to $11200$ cores in the tested systems to match the wall-clock time per time step of the whole DGX-2.} In regard to strong scaling of the GPU implementation on this system, one can depict in the present setup a small performance loss for the best-performing computational grids, between $5\%$ and $10\%$. This suggests that, as data is partitioned into smaller subdomains, the local problem size for this setup may become too small to fill the GPU efficiently.
\begin{table}[]
\centering
\begin{tabular}{|c|c|c|c|ll}
\cline{1-4}
\# GPU & \multicolumn{1}{l|}{grid} & Implicit diffusion off & Implicit diffusion on &  &  \\ \cline{1-4}
4      & $4\times 1$    &  $0.481\mathrm{s}$   &  $\mathbf{12.39}\mathrm{s}$    &  &  \\ \cline{1-4}
4      & $2\times 2$    &  $0.532\mathrm{s}$   &  $\mathbf{12.62}\mathrm{s}$    &  &  \\ \cline{1-4}
8      & $8\times 1$    &  $0.251\mathrm{s}$   &  $0.73\mathrm{s}$  &  &  \\ \cline{1-4}
8      & $4\times 2$    &  $0.275\mathrm{s}$   &  $0.846\mathrm{s}$ &  &  \\ \cline{1-4}
16     & $16\times 1$   &  $0.1404\mathrm{s}$  &  $0.398\mathrm{s}$ &  &  \\ \cline{1-4}
16     & $8\times 2$    &  $0.1477\mathrm{s}$  &  $0.444\mathrm{s}$ &  &  \\ \cline{1-4}
16     & $4\times 4$    &  $0.149\mathrm{s}$   &  $0.459\mathrm{s}$ &  &  \\ \cline{1-4}
\end{tabular}
    \caption{Wall-clock time per time step on a DGX-2 system. The numbers in bold are for runs that will not fit in the GPU memory. The timings of the same simulation using explicit diffusion on a $64\times 32$ computational grid, i.e.\ \new{$2048$} MPI tasks for the CPU simulation carried out on Beskow (Cray XC40, Xeon E5-2695v4/E5-2698v3 16C 2.3GHz, Aries interconnect) and Tetralith (Intel H2204XXLRE, Xeon Gold 6130 16C 2.1GHz, Intel Omni-Path) are respectively \new{$\mathbf{0.78}$ and $\mathbf{0.43}$} seconds per time step.}\label{tbl:performances}
\end{table}
\section{Conclusions and outlook}\label{sec:conclusions}
We have extended the open-source code \emph{CaNS} for massively-parallel simulations in GPU-accelerated systems. Since the original version of the code was implemented in Fortran90/95, CUDA Fortran is used for porting the code to GPUs with a manageable effort, while still achieving very good computational performance. The portability with CUDA Fortran has been further simplified by the novel, unified memory model which allows the programmer to define arrays that can reside on the host or on the device, without duplicating the arrays in the source code.\par
The implementation has been validated against benchmark data for turbulent channel flow, and the performance on a NVIDIA DGX-2 system has been examined. Sufficient data partitioning to ensure that the data resides mostly on the GPU is a key element for achieving a good performance, as excessive data migration between the CPU and GPU can severely degrade the performance. The results show that, remarkably, wall-clock time per time step using only $4$ of the $16$ GPUs of the DGX-2 system, is \new{just
slightly larger than the CPU implementation in \new{$2048$} cores of a state-of-the-art supercomputer, and about \new{$3$} faster times when the entire DGX-2 system is used.}\par
\new{Based on its good computational performance, and in particular the very low wall-clock time per time step, we believe that the current tool will serve well as a base Navier-Stokes solver on top of which numerical methods for simulating more complex phenomena on many-GPU systems (e.g.\ multiple phases and complex geometries) can be implemented.}\par
Both implementations are freely-available and open-source on GitHub, under the terms of an MIT license. See \texttt{\href{https://github.com/p-costa/CaNS}{github.com/p-costa/CaNS}}  for the original MPI-OpenMP implementation \cite{Costa-CAMWA-2018}, and \texttt{\href{https://github.com/maxcuda/CaNS}{github.com/maxcuda/CaNS}} for the implementation addressed in this manuscript.
\section*{Acknowledgments}
PC and LB acknowledge the funding from the European Research Council grant no.\ ERC-2013-CoG-616186, TRITOS, and the computing time provided by SNIC (Swedish National Infrastructure for Computing). PC acknowledges funding from the University of Iceland Recruitment Fund grant no.\ 1515-151341, TURBBLY. Finally, PC thanks Ali Yousefi from KTH Mechanics for producing the visualization in figure~\ref{fig:visu}.
%
%\section*{References}
\bibliography{bibfile}

\begin{thebibliography}{10}
\expandafter\ifx\csname url\endcsname\relax
  \def\url#1{\texttt{#1}}\fi
\expandafter\ifx\csname urlprefix\endcsname\relax\def\urlprefix{URL }\fi
\expandafter\ifx\csname href\endcsname\relax
  \def\href#1#2{#2} \def\path#1{#1}\fi

\bibitem{Pope-2000}
S.~B. Pope, Turbulent Flows, Cambridge University Press, 2000.
\newblock \href {http://dx.doi.org/10.1017/CBO9780511840531}
  {\path{doi:10.1017/CBO9780511840531}}.

\bibitem{Ishihara-et-al-ARFM-2009}
T.~Ishihara, T.~Gotoh, Y.~Kaneda, Study of high--reynolds number isotropic
  turbulence by direct numerical simulation, Annual Review of Fluid Mechanics
  41 (2009) 165--180.

\bibitem{Orszag-and-Patterson-PRL-1972}
S.~A. Orszag, G.~Patterson~Jr, Numerical simulation of three-dimensional
  homogeneous isotropic turbulence, Physical Review Letters 28~(2) (1972) 76.

\bibitem{top500}
\burl {https://www.top500.org/lists/2019/06/}, accessed on september 28, 2019.

\bibitem{MAGMA}
E.~Agullo, J.~Demmel, J.~Dongarra, B.~Hadri, J.~Kurzak, J.~Langou, H.~Ltaief,
  P.~Luszczek, S.~Tomov, Numerical linear algebra on emerging architectures:
  The plasma and magma projects, in: Journal of Physics: Conference Series,
  Vol. 180, IOP Publishing, 2009, p. 012037.

\bibitem{AFID-GPU}
X.~Zhu, E.~Phillips, V.~Spandan, J.~Donners, G.~Ruetsch, J.~Romero,
  R.~Ostilla-Monico, Y.~Yang, D.~Lohse, R.~Verzicco, M.~Fatica, R.~J. A.~M.
  Stevens, Afid-gpu: A versatile navier stokes solver for wall-bounded
  turbulent flows on gpu clusters, Computer Physics Communications 229.

\bibitem{Ha-et-al-JCP-2018}
S.~Ha, J.~Park, D.~You, A gpu-accelerated semi-implicit fractional-step method
  for numerical solutions of incompressible navier--stokes equations, Journal
  of Computational Physics 352 (2018) 246--264.

\bibitem{Ha-et-al-arXiv-2018}
S.~Ha, J.~Park, D.~You, A scalable multi-gpu method for semi-implicit
  fractional-step integration of incompressible navier-stokes equations, arXiv
  preprint arXiv:1812.01178.

\bibitem{Alfonsi-et-al-CaF-2014}
G.~Alfonsi, S.~A. Ciliberti, M.~Mancini, L.~Primavera, Gpgpu implementation of
  mixed spectral-finite difference computational code for the numerical
  integration of the three-dimensional time-dependent incompressible
  navier--stokes equations, Computers \& Fluids 102 (2014) 237--249.

\bibitem{Niemeyer-et-al-JS-2014}
K.~E. Niemeyer, C.-J. Sung, Recent progress and challenges in exploiting
  graphics processors in computational fluid dynamics, The Journal of
  Supercomputing 67~(2) (2014) 528--564.

\bibitem{Costa-CAMWA-2018}
P.~Costa, A fft-based finite-difference solver for massively-parallel direct
  numerical simulations of turbulent flows, Computers \& Mathematics with
  Applications 76~(8) (2018) 1853--1862.

\bibitem{Costa-et-al-PRL-2016}
P.~Costa, F.~Picano, L.~Brandt, W.-P. Breugem, Universal scaling laws for dense
  particle suspensions in turbulent wall-bounded flows, Physical review letters
  117~(13) (2016) 134501.

\bibitem{Dodd-and-Ferrante-JFM-2016}
M.~S. Dodd, A.~Ferrante, On the interaction of taylor length scale size
  droplets and isotropic turbulence, Journal of Fluid Mechanics 806 (2016)
  356--412.

\bibitem{Ostilla-et-al-JFM-2016}
R.~Ostilla-M{\'o}nico, R.~Verzicco, S.~Grossmann, D.~Lohse, The near-wall
  region of highly turbulent taylor--couette flow, Journal of fluid mechanics
  788 (2016) 95--117.

\bibitem{Wesseling-2009}
P.~Wesseling, Principles of computational fluid dynamics, Vol.~29, Springer
  Science \& Business Media, 2009.

\bibitem{Uhlmann-JCP-2005}
M.~Uhlmann, An immersed boundary method with direct forcing for the simulation
  of particulate flows, Journal of Computational Physics 209~(2) (2005)
  448--476.

\bibitem{Schumann-and-Sweet-JCP-1988}
U.~Schumann, R.~A. Sweet, Fast fourier transforms for direct solution of
  poisson's equation with staggered boundary conditions, Journal of
  Computational Physics 75~(1) (1988) 123--137.

\bibitem{Kim-and-Moin-JCP-1985}
J.~Kim, P.~Moin, Application of a fractional-step method to incompressible
  navier-stokes equations, Journal of computational physics 59~(2) (1985)
  308--323.

\bibitem{Frigo-and-Johnson-1998}
M.~Frigo, S.~G. Johnson, Fftw: An adaptive software architecture for the fft,
  in: Proceedings of the 1998 IEEE International Conference on Acoustics,
  Speech and Signal Processing, ICASSP'98 (Cat. No. 98CH36181), Vol.~3, IEEE,
  1998, pp. 1381--1384.

\bibitem{CUDAFortranBook}
G.~Ruetsch, M.~Fatica, {{CUDA}} Fortran for Scientists and Engineers, Morgan
  Kaufmann, 2013.

\bibitem{Makhoul-1980}
J.~Makhoul, A fast cosine transform in one and two dimensions, IEEE
  Transactions on Acoustics Speech and Signal Processing 28~(1) (1980) 27--34.

\bibitem{Hasbestan-and-Senocak-arXiv-2019}
J.~J. Hasbestan, I.~Senocak, Pittpack: An open-source poisson's equation solver
  for extreme-scale computing with accelerators, arXiv preprint
  arXiv:1909.05423.

\bibitem{NVTX}
\burl
  {http://devblogs.nvidia.com/parallelforall/cuda-pro-tip-generate-custom-application-profile-timelines-nvtx},
  accessed on august 21, 2019.

\bibitem{NVTXFortran}
\burl
  {https://devblogs.nvidia.com/parallelforall/customize-cuda-fortran-profiling-nvtx},
  accessed on august 21, 2019.

\bibitem{Moser99}
R.~D. Moser, J.~Kim, N.~N. Mansour, Direct numerical simulation of turbulent
  channel flow up to $\mathrm{Re}_\tau=590$, Physics of Fluids 11~(4) (1999)
  943--945.

\bibitem{Orlandi-2012}
P.~Orlandi, Fluid flow phenomena: a numerical toolkit, Vol.~55, Springer
  Science \& Business Media, 2012.

\bibitem{Henningson-and-Kim-JFM-1991}
D.~S. Henningson, J.~Kim, On turbulent spots in plane poiseuille flow, Journal
  of Fluid Mechanics 228 (1991) 183--205.

\end{thebibliography}
\end{document}